# EMPOWERING HEALTH IN AGING: INNOVATION IN UNDERNUTRITION DETECTION AND PREVENTION THROUGH COMPREHENSIVE MONITORING

Abderrahim Derouiche[1], Ghazi Bouaziz[1], Damien Brulin[1],
Eric Campo[1] and Antoine Piau[2]
[1]*LAAS-CNRS, University of Toulouse, CNRS, UPS, UT2J, Toulouse, France*
[2]*Center for Epidemiology and Population Health Research, CHU Toulouse, Toulouse, France*

**ABSTRACT**

Addressing the health challenges faced by the aging population, particularly undernutrition, is of paramount importance, given the significant representation of older individuals in society. Undernutrition arises from a disbalance between nutritional intake and energy expenditure, making its diagnosis crucial. Advances in technology allowed a better precision and efficiency of biomarker measurements, making it easier to detect undernutrition in the elderly. This article introduces an innovative system developed as part of the CART initiative at Toulouse University Hospital in France. This system takes a comprehensive approach to monitor health and well-being, collecting data that can provide insights, shape health outcomes, and even predict them. A key focus of this system is on identifying nutrition-related behaviors. By integrating quantitative and clinical assessments, which include biannual nutritional evaluations, as well as physical and physiological measurements like mobility and weight, this approach improves the diagnosis and prevention of undernutrition risks. It offers a more holistic perspective aligned with physiological standards. An example is given with the data collection of an elderly person followed at home for 3 months. We believe that this advance could make a significant contribution to the overall improvement of health and well-being, particularly in the elderly population.

## 1. INTRODUCTION

People aged 65 and over living alone represent a considerable proportion of the population in the US (13%) and France (19.6%) (Perez et al., 2021). However, monitoring undernutrition in the elderly has been little studied using technological devices. It is based on medical assessments using weighing, directed or semi-directed questionnaires (Reber et al., 2019). Undernutrition is a pathological condition resulting from insufficient nutritional intake in relation to the body's energy expenditure. It is a major public health issue in the management of older people with chronic pathologies (up to 50% of elderly people) (Volkert et al., 2019). Indeed, it is associated with increased morbidity and mortality, risk of loss of autonomy, falls, infections and emergency hospitalizations (Norman, Haß and Pirlich, 2021). In France, 400 000 elderly living at home suffer from undernutrition, so it is important to better understand the behaviors related to undernutrition. The use of digital biomarkers for the early detection of functional or cognitive decline is receiving increasing attention (Piau et al., 2020). For example, sensor-derived mobility indicators have been studied to predict an upcoming fall in older people (Schwenk et al., 2014). Currently, the assessment of an individual's eating behavior relies on self-report measures and clinical tests. Although this methodology is necessary as a first step, it can be improved as it is limited by recall bias and by brief and mostly episodic assessments. Continuous real-time monitoring by a low-cost ambient sensors network can provide a more sensitive, objective and environmentally valid method to:

- Detect sub-clinical alteration in nutrition during chronic disease, prior to the onset of symptoms (allowing for targeted, personalized and more effective medicine, rather than one-size-fits-all medicine).

- Make a less biased assessment of treatment efficacy and side effects in a clinical trial after treatment modification.

- Remotely monitor the dynamics of recovery after hospitalization or surgery.

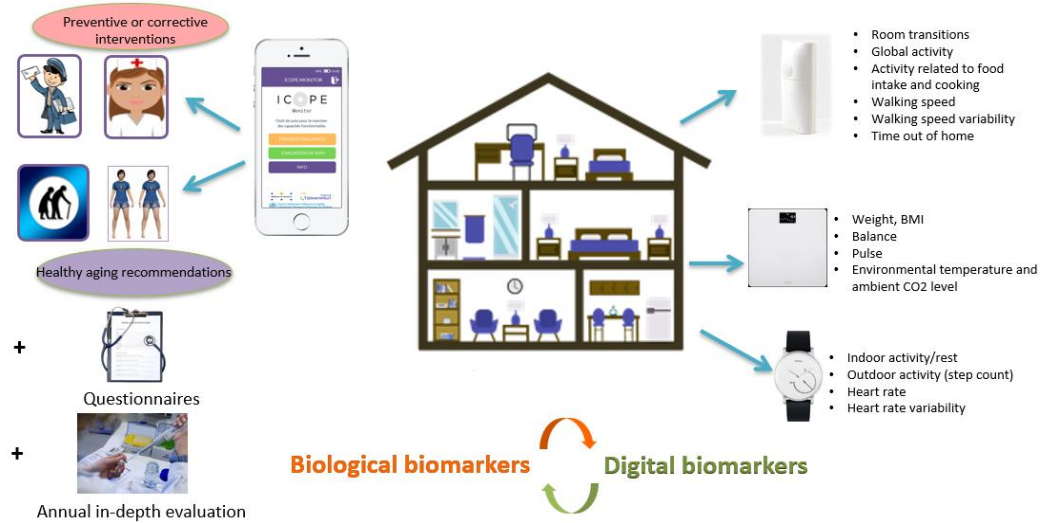

Figure 1. CART France (Derouiche et al., 2023)

As shown in Figure 1, the aim of the CART project, in which this work is involved, is to meet the constraints of digital analysis with a complementary clinical study. Clinical analysis includes a self-assessment examination to identify biological biomarkers, followed by their correlation with digital biomarkers by a clinical expert. In our study, we have implemented a part of this system in the homes of several elderly people living alone. This article presents a non-exhaustive literature review of related works in section 2. Section 3 aims to describe our system architecture. Section 4 presents some results obtained with an elderly person monitored for 3 months at home, followed by a discussion. Finally, the paper ends with a conclusion and some perspectives.

## 2. RELATED WORKS

Previous studies have shown distinct links between the physiological state and the biological state of a person, providing a strong clue to the detection of undernutrition (Ferguson et al., 2016).

Busnel and Ludwig (2018) proposed to estimate the diagnosis accuracy of the nutritional status score, body mass index and weight loss documented from the Resident Assessment Instrument – Home Care (RAI-HC). A sample of 267 home care recipients aged 65 years and older was assessed using the RAI-HC and the Mini Nutritional Assessment Short Form (MNA-SF). The result reveals that the diagnostic accuracy of the RAI-HC indicators was not sufficient for optimal screening regarding undernutrition in elderly home care recipients. Additional assessment with the MNA-SF is recommended to optimize early detection of individuals at risk for undernutrition.

İlhan and colleagues (2018) proposed to investigate the reliability and validity of the Turkish version of the Simplified Nutritional Appetite Questionnaire (SNAQ) in geriatric outpatients. Cohen's kappa analysis showed fair to moderate agreement between the SNAQ and the MNA ($\kappa=0.355$, $p<0.001$). Female gender, illiteracy, functional dependence in activities of daily living was significantly associated with poor appetite. The SNAQ score was weakly correlated with the MNA-SF and MNA-LF scores ($r=0.392$ and $r=0.380$, respectively, $p<0.0001$ for both).

Gambi, Ricciuti and De Santis (2020) proposed an application for monitoring an individual's eating habits during a meal, with the eating actions being assessed through in-depth image analysis. To enhance algorithmic performance and efficiency, the author's concept involves automating the process to identify various food-related actions.

Hilde Lohne-Seiler and others (2014) introduced the ActiGraph GT1M accelerometer, which was used to measure Physical Activity (PA) over seven consecutive days. Additionally, a questionnaire was used to document self-reported health status. PA levels among Norwegian seniors varied with age. In general, the elderly population spends 66% of their time in a sedentary state, with only 3% engaging in moderate-to-vigorous physical activity (MVPA). Twenty-one percent of participants met the current Norwegian PA recommendations. Furthermore, PA levels were correlated with self-reported health status.

Minh Pham and colleagues (2016) introduced a new concept involving the use of a smart home. This framework consists of a 16x22-foot experimental area equipped with environmental sensors, home service robots, a human subject equipped with wearable sensors and mobile devices such as smartphones. All these components are interconnected by a home gateway. The model was seamlessly integrated into this gateway to categorize daily activities (such as sitting, standing, lying down, walking, going up and down stairs, jogging and running). The resulting labels were transmitted continuously and in real time to the cloud for monitoring purposes. Meanwhile, raw data, time-stamped acceleration, and angular rates, were uploaded to the cloud once every 24 hours, where they were stored for later analysis. The authors wanted to demonstrate the potential of their system by creating a hydration monitoring application. After extensive testing, the system achieved a remarkable 91.5% accuracy rate in classifying drinking activities using k-fold cross-validation. But the problem lies in the limited amount of physiological data available.

Hoca and Turker (2021) assessed Body Mass Index (BMI) and muscle strength using a hand dynamometer. Elderly people with a BMI $>= 30$ kg/m$^2$ had significantly lower grip strength for the right and left hand than those with a BMI between 18.5 and 24.9 kg/m$^2$ and between 25 and 29.9 kg/m$^2$. However, those with a BMI $>= 30$ kg/m$^2$ had significantly higher values for waist/hip ratio, waist/height ratio, body fat percentage, waist circumference, hip circumference, neck circumference, neck circumference and mid-upper arm circumference compared to those with a BMI of 18.5-24.9 kg/m$^2$ and 25-29.9 kg/m$^2$. In addition, the authors observed a positive and statistically significant correlation between right- and left-hand grip strength and MNA score in women.

One of the main findings of previous studies is that weight loss is a fundamental marker. Other studies also suggest that albumin levels in laboratory tests may be related to undernutrition. Finally, the identification of biomarkers (Panagoulias, Sotiropoulos and Tsihrintzis, 2021) is also proving to be a tool for accurately prognosticating many diseases, classifying them by type and stage, and personalizing medical care.

Weight loss can be monitored by regular weighing. However, the person needs to be able to weigh themselves periodically and report their weight to attending physician. Another approach is to monitor mobility parameters while estimating energy expenditure based on the individual's self-reported daily routines and meal-related activities in the kitchen (McClung et al., 2018).

## 3. PROPOSED SYSTEM

The proposed research is part of the regional Inspire program (De Souto Barreto et al., 2020). Inspire is a research platform dedicated to gerontological research on biological and healthy ageing to identify biomarkers, determine biological age and work on all marks of ageing. Inspire involves one thousand participants. The two main projects are ICOPE and CART.

### 3.1 ICOPE Mobile Application

As shown in Figure 2, Integrated Care for Older People (ICOPE) (Tavassoli et al., 2021) is a mobile application launched by the World Health Organization (WHO) in 2019 for the integrated care of older people. This app measures intrinsic capacity which is the combination of different abilities (physical, mental and psychological) and functional ability. There are six domains of intrinsic capacity that can be measured: cognition, mobility, nutrition, mood, vision and hearing. To maintain the health of older people and to better identify risks before the frailty stage is reached, the application is used as a monitoring tool. It is based on five steps: 1 - Screening for decline in intrinsic capacity, 2 - Specialized assessment, 3 - Creating a personal care plan, 4 - Regular monitoring and 5 - Career and community integration. To monitor undernutrition, patients answered two questions about:
 - Weight loss: have you unintentionally lost more than 3 kg over the last three months?
 - Appetite loss: have you experienced a loss of appetite?

By answering these two questions, we can identify markers related to nutrition. We can also establish a link between nutrition and reduced mobility.

To better collect biological markers, an annual in-depth assessment and questionnaires is used.

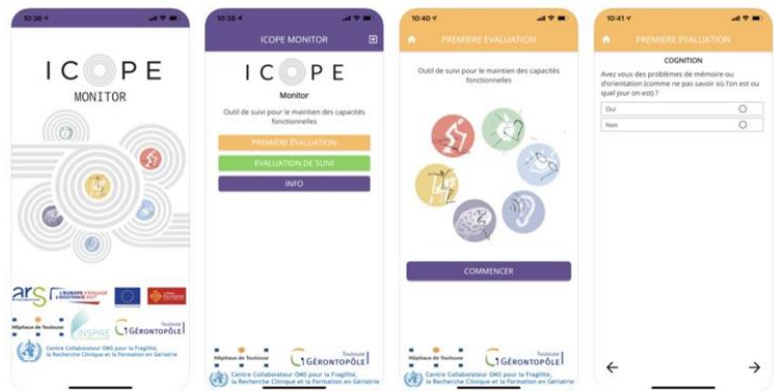

Figure 2. ICOPE Mobile application

## 3.2 CART France

Collaborative Aging Research using Technology (CART) Sensor Platform is an initiative developed by Oregon Center for Aging & Technology - ORCATECH to study various physical and physiological parameters (Thomas et al., 2021). CART France is a living lab project based on one hundred connected homes. ORCATECH's sensor platform proposes new ways to assess people with cognitive impairment. By using a huge amount of information about changes in health status, activity and functioning in a real-world setting, one can improve the ability to assess and provide personalized care. Differences between the French CART project and the U.S. CART project are the number of sensors used in both projects and the fact that the French CART project is focused on the nutritional status of individuals. As shown in Figure 3, the sensors used in CART France are physical and physiological sensors.

The project aims to combine the CART French data with clinical data to evaluate the relevance of numerical data to represent and predict an individual's health status.

The inclusion criteria for participants are as follows:
1- Over 75 years old
2- Living alone at home without assistance
3- Being robust
4- Having a computer and an Internet connection

To retrieve the physical parameters, we used two types of passive infrared motion sensors and contact sensors installed at 3 distinct locations in the individual's home:

- the 1st type of motion sensor is a wall sensor. It is used to detect the presence of the person in the room (one per monitored room).

- the 2nd type of motion sensor is a line sensor; 4 of them are placed in the hallway to calculate the walking speed when the person moves between two rooms. This parameter has already been studied to predict falls (Piau et al., 2020).

- the 3rd sensor is a contact sensor installed on the front door to detect the entry and exit of the person from the house.

The sensors are wireless and use ZigBee radio communication. To retrieve physiological parameters, other types of sensors such as an impedance scale and a connected smart watch are added. Detailed descriptions of the hub computer and sensors have been published previously (Beattie et al., 2020) (Thomas et al., 2020).

Finally, the CART France project (Derouiche et al., 2023) signifies significant research undertaking with the capacity to greatly impact the advancement of healthy and independent aging. Nevertheless, it is crucial to comprehensively tackle ethical considerations throughout the project, ensuring the safeguarding of participants' rights and welfare.

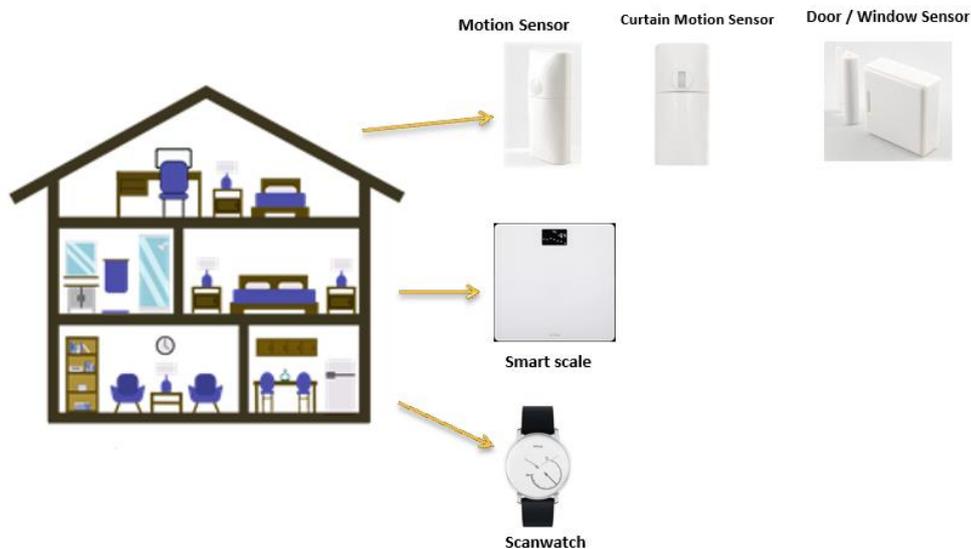

Figure 3. Overview of devices in the CART France project

## 3.3 Ethics and User Acceptance

French data Protection Authority (DPA) and Committee for the Protection of Persons concerned (CPP) approvals have been obtained before the beginning of installation phase. The introduction of passive technologies, including in-room passive sensors and activity-tracking sensors was unanimously approved by all participants. Most technologies operating unobtrusively, such as wall and line sensors and window and door sensors, were well-received in the opinion of residents. Emphasizing the importance of upholding residents' rights and preserving their dignity, participants reached a consensus on the need to protect personal and medical information, ensuring that it remains confidential and is not shared with third parties without their explicit consent.

## 3.4 System Architecture

The French CART project was launched in July 2022. In this section we will present one of the first installations using only presence sensors and a contact sensor to monitor physical activities. The positions of the sensors in the house are shown in Figure 4.

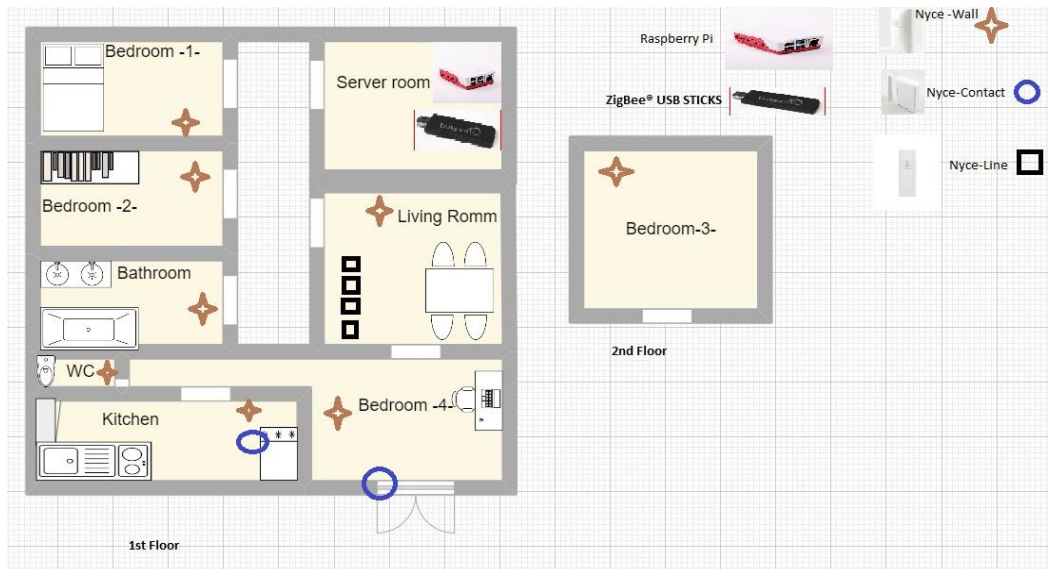

Figure 4. Typical house architecture

The exact number of each type of sensor required (in a total set of 13 sensors) is:
- 04 Line sensors in the living room;
- 08 Wall sensors (one per room).
- 02 Contact sensors.

Figure 5 shows the location of sensors in the home of a 77-year-old. Since the installation is done, we have been able to check, using a Raspberry Pi (RPi) control panel that all the sensors are working correctly, as shown in Figure 6. These sensors are connected to a hub (Rpi) via a personal area network (PAN).

The critical step involves establishing the positioning of sensors within the household. This entails providing precise data for each room, as depicted in Figure 7. Utilizing the sensor map not only helps us pinpoint sensor locations but also provides accurate data regarding the specific sensor placement. This, in turn, enables us to ascertain the appropriate type of indicator required for each location.

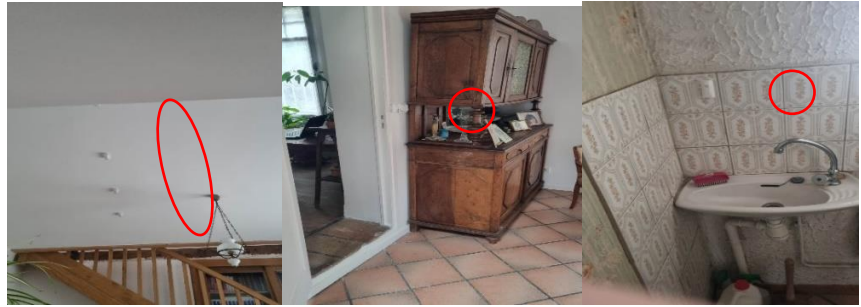

Figure 5. Example of sensors installation in a house

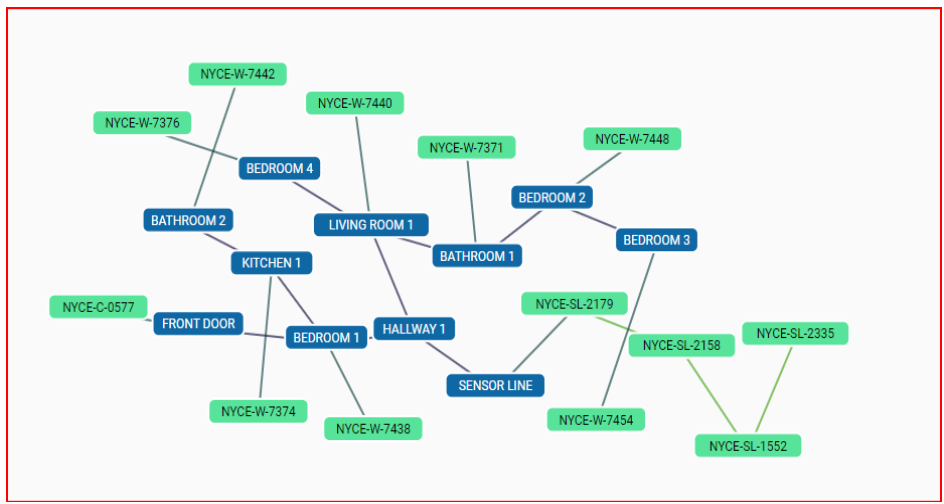

Figure 6. Example of sensors assigned in the console

Figure 7. Location of sensors on the console

## 3.5 Data Collection and Analysis

Up until now, we have collected 12 months of data. The objective is to determine the life pattern of this person by correlating these data with the medical data from the weekly questionnaire and the ICOPE application. The collection and analysis of declarative data from the ICOPE mobile application are conducted in parallel to our study by the Hospital. The questionnaire covers mobility, sight, hearing, memory, nutritional status and morale and will be useful for the specialist to merge with digital biomarkers.

The analysis described here is based on a naive approach that does not consider noise factors such as early triggering of sensors, transmission delay, or missing events, but it is sufficient to give a basic idea of their functionality. The data is stored in a database that includes all data collected from all sensors. The obtained dataset is used to identify any walking speed during movement, based on the activation sequences of the four ceiling-mounted motion sensors (1, 2, 3, 4 or 4, 3, 2, 1). These activations were arranged in chronological order based on timestamps, ensuring that all sensors were activated. Figure 8 shows the event logs in the database including homeid, stamp (date&time), areaid (room), itemid (sensors) and event (Wall or sensor line sensors, 0: No Presence | 1: Presence. Contact (door) sensors, 0: Closed | 1: Open.).

Figure 8. Data Collection with area identification

Figure 9. First Pre-Processing

Then, we only keep the information necessary for the analysis (stamp, areaid, events). On the other hand, to make correct and fast calculations, data filtering becomes highly beneficial when working with Unix time to represent time intervals. Employing Unix time as the timestamp format streamlines the task of comparing and filtering events, facilitating smooth operations based on their temporal attributes as shown in Figure 10.

Figure 10. Filtering of all events by areaid

Given the inter-individual variability in the physical activity intensity and duration, total energy expenditure can be estimated by multiplying resting energy expenditure by a factor reflecting the intensity of an individual's physical activity. This physical activity level factor (PAL) has been determined for many activities of daily living, including sedentary, occupational and sports activities (Westerterp, 2013).

$$PAL = \frac{TotalEnergyExpenditure}{BaselMetabolicRate} \quad (1)$$

The Basel Metabolism Rate (BMR) is obtained by the following basal metabolic rate formula:
- Men: BMR=88.362+ (13,397×weight in kg) + (4,799×height in cm) – (5,677×age in years)
- Women: BMR=447.593+ (9,247×weight in kg) + (3,098×height in cm) – (4,330×age in years)

$$\text{Total energy expenditure}=PAL \times BMR \quad (2)$$

Establishing predefined rules for food-related activities is an essential step in understanding and analyzing the nutritional behaviors of individuals. These rules make it possible to categorize and identify specific activities and behaviors associated with food consumption and preparation. A sequence of events or conditions that corresponds to a food-related activity is established:
- It begins when a person enters the kitchen or opens the refrigerator or cupboard.
- If the person opens the fridge or cupboard after entering the kitchen, we consider this as a part of the eating activity.
- The activity ends when the individual is no longer in the kitchen or dining room.

Once these rules have been applied to our dataset, we can calculate correlations and analyze how different rooms and mobility patterns are related to eating-related activities (Figure 11). Steps to create our correlation matrix:
- **Create a New Variable:** Start by adding a new category in the dataset to represent activities related to nutrition. This variable will be assigned values based on specific criteria, using a systematic approach.
- **Organize the Data:** Once the nutrition-related variable is established, we organize the data by selecting the essential columns needed for calculating presence, occupancy rates, and mobility. These columns typically include "areaid," "event," "timestep," and the newly created variable for nutrition-related activities

- **Calculate Presence and Occupancy:** we Determine how often different rooms (such as the kitchen, dining pantry, fridge, and dining room) are used for nutrition-related activities. This is done by grouping the data based on the room's identification ("areaid") and finding the average frequency of events in each room.
- **Assess Mobility:** Evaluate how mobile individuals are within each room of the house. This is achieved we use the same grouping method as in the previous step, calculating the average level of activity within each room.
- **Create a Correlation Matrix:** Build a correlation matrix to visually represent the relationships between presence, occupancy rates, and mobility in each room. This matrix helps identify patterns and connections among these factors.

The process is carried out monthly in order to perform a more comprehensive analysis and monitor trends over a longer period. This means that the steps described are carried out at regular intervals. This consistent frequency will provide an overview of the progression of nutrition-related activities, presence, occupancy, and mobility over time.

By analyzing the correlation matrix, we will be able to determine which rooms are most closely associated with nutrition-related activities and how the mobility of individuals at home is linked to these activities.

The correlation matrix between presence and occupancy rates and mobility for each room can be calculated using Pearson's correlation formula (Sedgwick, 2012).

$$\text{Pearson Correlation}(A, B) = \frac{\sum_{i=1}^{n}(A_i - \overline{A})(B_i - \overline{B})}{\sqrt{\sum_{i=1}^{n}(A_i - \overline{A})^2 \quad \sum_{i=1}^{n}(B_i - \overline{B})^2}} \quad (3)$$

$A_i, B_i$ : A and B values for observation i.
$\overline{A}, \overline{B}$ : the means of variables A and B, respectively.
n is the total number of observations.

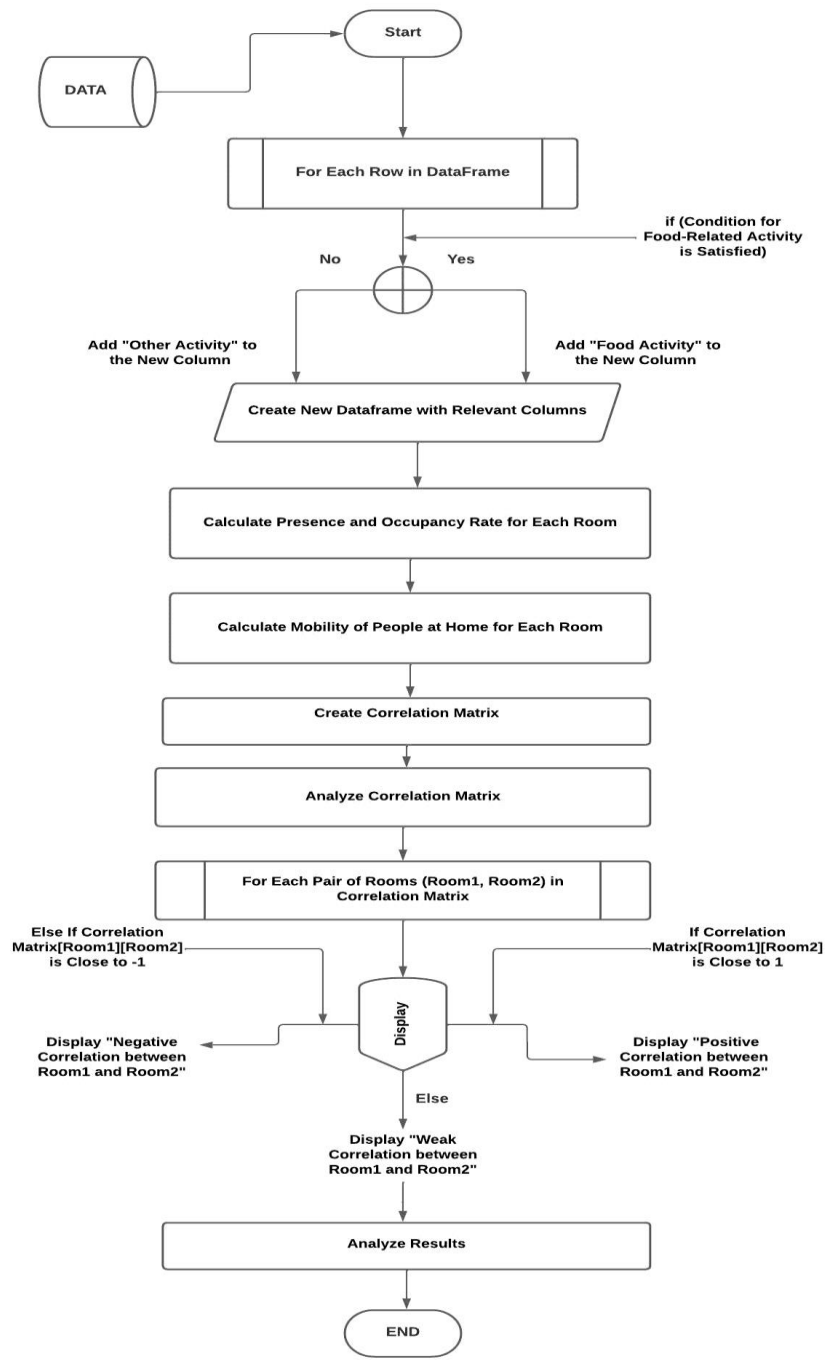

Figure 11. Flowchart of the correlation matrix creation and calculation process

EMPOWERING HEALTH IN AGING: INNOVATION IN UNDERNUTRITION DETECTION
AND PREVENTION THROUGH COMPREHENSIVE MONITORING

## 4. RESULTS AND DISCUSSION

Figure 12 shows an example of different events in each room of the house for 3 months of data for the subject selected in this paper. To better illustrate each of them, we represent each room by a histogram as shown in Figure 13. Each room contains events from which we can well distinguish the different activities of daily living (Mlinac and Feng, 2016), thus identifying a life pattern of that person. Finally, this will give us the ability to identify personalized digital biomarkers.

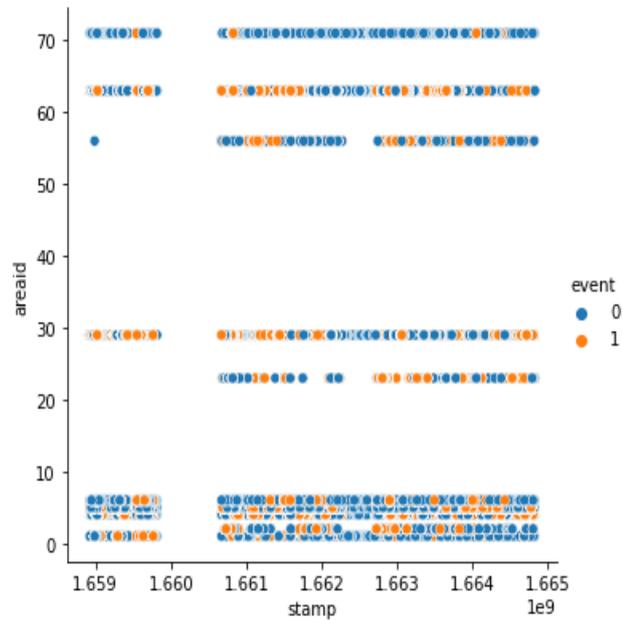

| Areaid | Area |
|---|---|
| 1 | Bathroom 1 |
| 2 | Bathroom 2 |
| 3 | Bathroom 3 |
| 4 | Badroom 1 |
| 5 | Badroom 2 |
| 23 | Kitchen |
| 29 | Living Room1 |
| 35 | Dining Closet |
| 56 | Front Door |
| 58 | Refregirator |
| 63 | Hallway |
| 71 | Bedroom 4 |

Figure 12. All the events that occurred in the house for 3 months

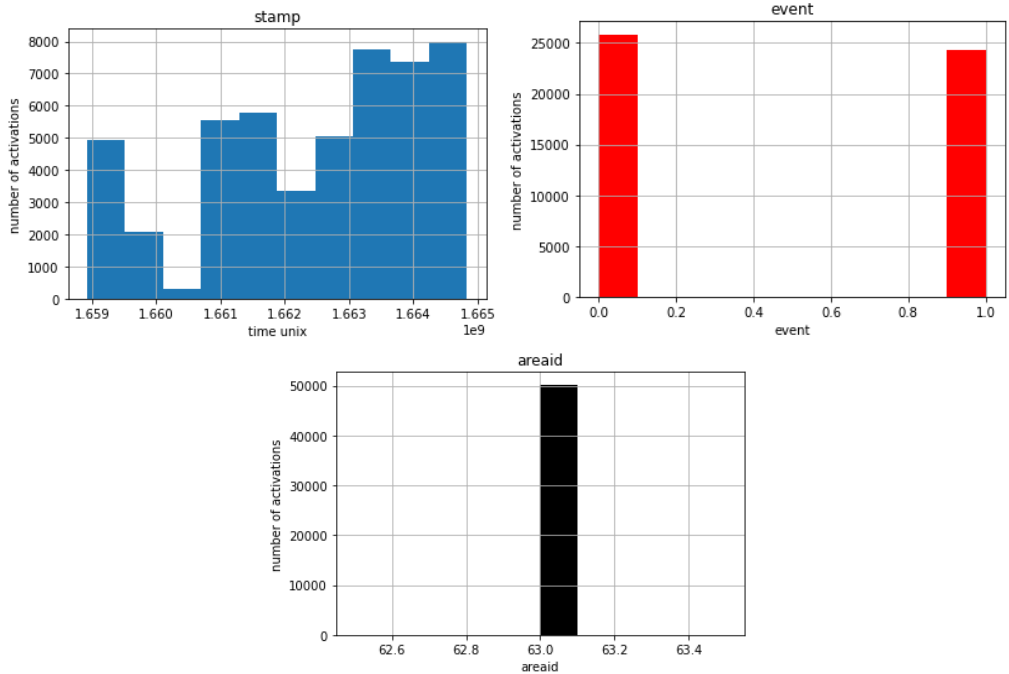

Figure 13. All events that occurred in area 63 (Hallway) for 3 months

We can also conclude that the use of infrared motion sensors not only detect the various activities of the person but it can also show the lifestyle which we can, associated to other parameters, allow to develop a specific model for that person. It can also be used to calculate caloric expenditure. Considering the following characteristics (age 77 years, sex: male, height 165 cm, weight 60 kg and work: writer), the BMR of this subject is:

$$BMR_{Men} = 88.362 + (13.397 \times 60) + (4.7988 \times 165) - (5,677 \times 77) = 1246.855 \text{ Cal.}$$

Based on the lifestyle of the person in the house, we can classify this person in the PAL category between [1.4 and 1.69], which means he has a low-activity lifestyle, we typically retain only the lower bounds of the Physical Activity Level (PAL) range to simplify calculations and obtain a more conservative estimate of energy expenditure. The lower bound of PAL represents the minimum level of activity required to maintain basic bodily functions, such as respiration and digestion, in the absence of additional physical activity. The equation (2) gives:

$$Total\ Energy\ Expenditure = 1.4 \times 1\ 246.855\ cal = 1745.597 \text{ Cal.}$$

From the results of the weekly questionnaire, we can theoretically identify the lifestyle of the participants and validate this study by cross-referencing the activity records. To enhance the visualization of nutritional status, one strategy involves expanding the sensor network by integrating contact sensors into commonly used kitchen fixtures, such as dining cupboards and microwaves. We focus primarily on nutrition-related activities, which lead us to target areas associated with dietary practices, including the eating cupboard, kitchen, fridge, and dining room, as shown in Figure 14. The objective is to enhance tracking capabilities by introducing supplementary sensors in the kitchen while upholding the existing level of comfort and without disrupting the routines of participants during the experiment, a feature already provided by the CART experiment.

However, the results of our analysis slightly approximate the amount of daily energy spent in the home only. If we wish to calculate total daily energy expenditure, we must include other types of sensors, such as a smartwatch or a smart scale, to monitor the total amount of activity inside and outside the home.

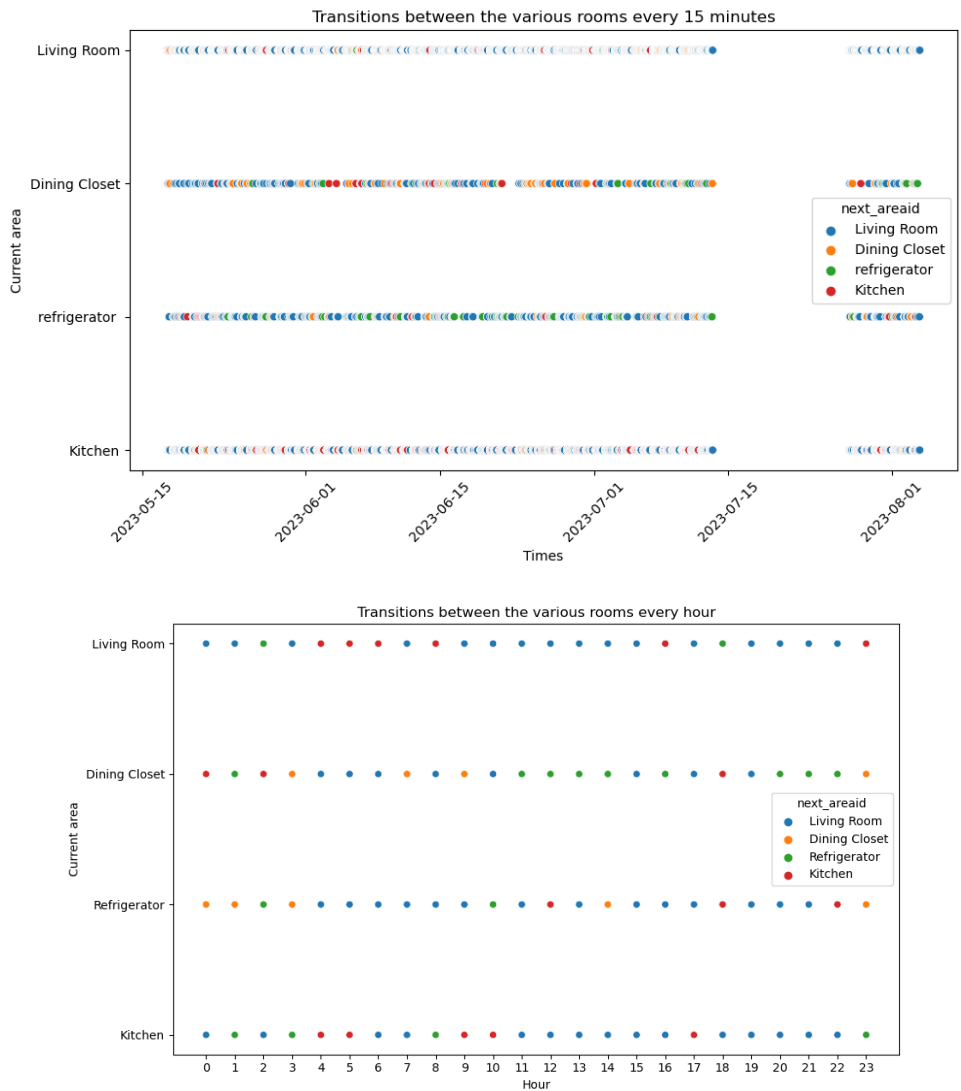

Figure 14. Transitions between rooms

Transitions between rooms allow us to consider additional parameters such as kitchen occupancy, occupancy rates and average room occupancy.

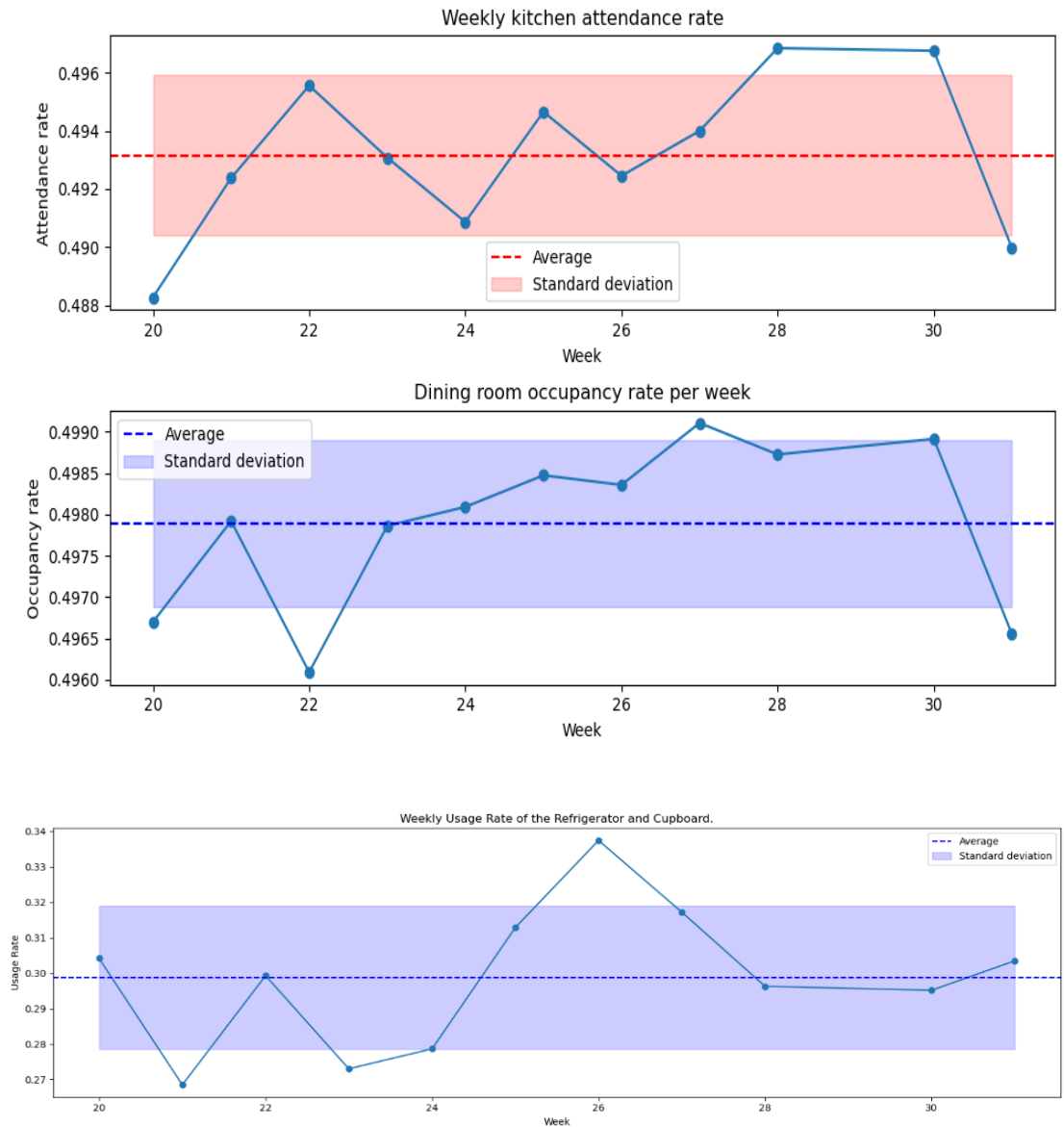

Figure 15. Presence and Occupancy Rates in Food-Related Activities

EMPOWERING HEALTH IN AGING: INNOVATION IN UNDERNUTRITION DETECTION
AND PREVENTION THROUGH COMPREHENSIVE MONITORING

Several indicators are valuable metrics for assessing physical activity, dietary habits, and in-home mobility, contributing to more effective health management and improved quality of life. They are commonly used in research, personalized healthcare, and home behavior analysis.

Those we have selected are:

1 - Total Energy Expenditure is frequently used in conjunction with mobility data to obtain a more comprehensive understanding of an individual's physical activity and energy requirements.

2 - Kitchen Presence and Occupancy Rates for Nutritional Activities This metric is used to comprehend and evaluate dietary behaviors. It involves monitoring the presence in the kitchen and the time spent in this area during nutrition-related activities, such as meal preparation.

3 - In-Home Mobility is a versatile metric with applications spanning healthcare, safety, personalized care, and research. Its value lies in its ability to provide deeper insights into how individuals move and interact within their home environment, ultimately leading to enhanced health management and quality of life.

Beyond the indicators, we have calculated a correlation matrix, focusing on the links between the different 'areaid'. This matrix allows us to identify correlations between different areas of the home. In particular, it can provide valuable information on eating-related activities and in-home mobility. For instance, with respect to kitchen indicators, our attention can be directed toward activities like meal preparation and the time spent in the kitchen. When it comes to in-home mobility, we can focus on activities such as transitioning between rooms and the associated behaviors .and on how nutrition-related activities and mobility interact, evolve, and influence occupants' habits over time. By using the correlation matrix in conjunction with mobility data, we aim to obtain a more comprehensive perspective on an individual's physical activity and energy requirements.

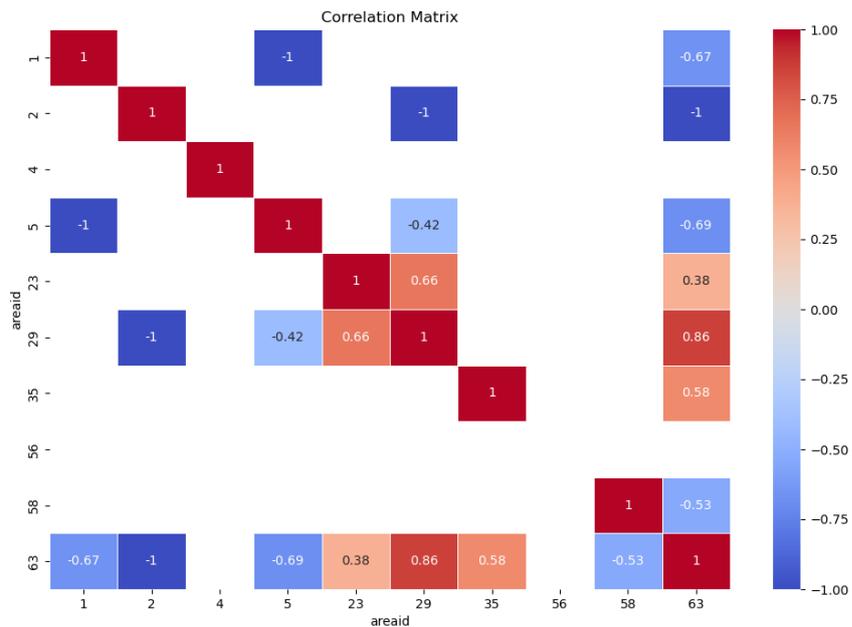

Figure 16. Correlation matrix between areas (areaid)

We have considered three distinct levels:

**1.** Values Close to 1 (Positive Correlation):

- A value close to 1 indicates a strong positive correlation.

- When activity in one room increases, activity in another room also increases, and vice versa.

- In the context of food-related activities, this could mean that some rooms are systematically used to prepare, eat, or store food.

**2.** Values Close to -1 (Negative Correlation):

- A value close to -1 indicates a strong negative correlation.

- When activity in one room increases, activity in another room decreases, and vice versa.

- In the context of food-related activities, this could mean that some rooms are used in opposition to food-related tasks, such as people avoiding certain rooms when cooking or eating.

**3.** Values Close to 0 (Low Correlation):

- A value close to 0 suggests a weak or no correlation.

- There is little or no relationship between activity in one room and activity in another.

- In the context of food-related activities, this may indicate that some rooms are relatively independent of feeding-related tasks.

In summary, by scrutinizing the correlations between specific rooms and activities linked to nutrition, we can ascertain which rooms display pronounced associations with these nutrition-related activities, whether positively or negatively. Additionally, we can identify rooms that maintain a lesser influence from nutrition-related tasks. These findings contribute to a deeper understanding of behavioral patterns related to nutrition within a particular household environment.

When these factors are combined, they define a novel digital biomarker. This biomarker can be harnessed as part of an alert system designed to detect subtle shifts in behavior, a validation process that greatly benefits from the insights of a healthcare professional.

## 5. CONCLUSION

In conclusion, our paper has introduced a new system developed as part of the CART initiative, from regional research projects at the Toulouse University Hospital in France. This system is designed to comprehensively monitor various aspects of health and well-being. It collects data to follow, detect and predict health status over the time, with an emphasis on identifying nutrition-related behaviors.

By gathering information on individuals' daily habits, including their presence in specific rooms like the kitchen and their use of household appliances such as the refrigerator, we have established the foundations for assessing nutritional behaviors. This foundation is reinforced by the crossing of three key indicators: total energy expenditure, presence and occupancy rates of rooms for nutritional activities, and the mobility of individuals at home. Analysis of the correlation matrix leads to several conclusions:

- Identify key rooms: Rooms with a strong positive correlation with food-related activities are likely to be at the heart of food preparation and consumption. These may include the kitchen or dining room.

- Identify avoidance habits: Rooms with a strong negative correlation may indicate areas where people tend to avoid food-related activities. For example, if the bedroom or home office show a strong negative correlation, this may indicate that people avoid eating or cooking in these areas.

- Understanding mobility: The correlation matrix can also reveal the link between people's mobility at home and food-related activities. If, for example, there is a strong positive correlation between activity in the kitchen and the dining room, this suggests that people move between these rooms when they eat.

Merging these indicators creates an innovative digital biomarker. The combination of digital data and clinical assessments, such as biannual nutritional evaluations and measurements of physical and physiological parameters, offers a promising approach to improving the diagnosis and prevention of situations at risk of undernutrition. This combined approach enables us to better understand individuals' adherence to recommendations and provides a global perspective closer to physiological conditions, thus advancing our ability to improve overall health and well-being.